\documentclass[letter]{aa}         
\usepackage{txfonts}
\usepackage{graphicx}
\bibpunct{(}{)}{;}{a}{}{,}      

\begin{document}

\title{Simulations of fully deformed oscillating flux tubes}

\author{K. Karampelas\inst{1} \and T. Van Doorsselaere\inst{1}} 

\institute{Centre for mathematical Plasma Astrophysics, Department of Mathematics, KU Leuven, Celestijnenlaan 200B bus 2400, 3001 Leuven, Belgium \\ \email{kostas.karampelas@kuleuven.be}} 

\date{Received <date> /
Accepted <date>}

\abstract{In recent years, a number of numerical studies have been focusing on the significance of the Kelvin-Helmholtz instability (KHI) in the dynamics of oscillating coronal loops. This process enhances the transfer of energy into smaller scales, and has been connected with heating of coronal loops, when dissipation mechanisms, such as resistivity, are considered. However, the turbulent layer is expected near the outer regions of the loops. Therefore, the effects of wave heating are expected to be confined to the loop's external layers, leaving their denser inner parts without a heating mechanism.}
{In the current work we aim to study the spatial evolution of wave heating effects from a footpoint driven standing kink wave in a coronal loop.}
{Using the MPI-AMRVAC code, we performed ideal, three dimensional magnetohydrodynamic simulations of footpoint driven transverse oscillations of a cold, straight coronal flux tube, embedded in a hotter environment. We have also constructed forward models for our simulation using the FoMo code.}
{The developed Transverse Wave Induced Kelvin-Helmholtz (TWIKH) rolls expand throughout the tube cross-section, and cover it entirely. This turbulence significantly alters the initial density profile, leading to a fully deformed cross section. As a consequence, the resistive and viscous heating rate both increase over the entire loop cross section. The resistive heating rate takes its maximum values near the footpoints, while the viscous heating rate at the apex.}
{We conclude that even a monoperiodic driver can spread wave heating over the whole loop cross section, potentially providing a heating source in the inner loop region. Despite the loop's fully deformed structure, forward modelling still shows the structure appearing as a loop.}

\keywords{magnetohydrodynamics (MHD) - Sun: corona - Sun: oscillations} 

\maketitle                      

\titlerunning{Simulations of fully deformed oscillating flux tubes}
\authorrunning{K. Karampelas et al.}

\section{introduction}
Heating of coronal loops by transverse magnetohydrodynamic (MHD) waves has been an extensively studied topic ever since the proof of their ubiquity in the solar atmosphere \citep{aschwanden1999, tomczyk2007}. The main theory of wave damping is resonant absorption for the case of standing modes \citep{Ionson1978ApJ, goossens1992resonant, arregui2005resonantly, terradas2010, goossens2011resonant} and its analogous mechanism of mode coupling \citep{pascoe2010, demoortel2016} for propagating waves. In both mechanisms the energy of the large-scale oscillation is transferred, through resonance, to local azimuthal Alfv\'{e}n modes. In the case in which multiple frequencies are excited, smaller scales are created through phase mixing \citep{heyvaerts1983,soler2015}. By including dissipation mechanisms, such as resistivity or viscosity, resonant absorption and mode coupling can lead to heating \citep{ofman1998, pagano2017}.

However, the effects of wave heating by global oscillations were believed to be confined in the resonant layer. As shown in \citet{cargill2016ApJ}, this localized heating is not capable of sustaining a fixed density gradient between the loop and the environment. Radiative cooling will inevitably lead to draining of the loop's denser inner parts, unless additional heating mechanisms are considered. A possible solution could be the use of a broad-band driver for transverse waves. In such a case \citep{ofman1998}, we would see the development of multiple narrow resonance layers. These layers can move across the loop cross-section as the density profile changes, but heating would still be concentrated in near-discrete locations.

The previous issue could also be potentially addressed by the Kelvin-Helmholtz Instability (KHI) for standing modes in closed coronal structures \citep{heyvaerts1983, zaqarashvili2015ApJ}. Its existence is predicted by three dimensional simulations in straight flux tubes for driver generated azimuthal Alfv\'{e}n waves \citep{ofman1994nonlinear,poedts1997b}, impulsively excited standing kink modes \citep{terradas2008,antolin2014fine,magyar2015,magyar2016damping,howson2017}, and footpoint driven standing kink modes \citep{karampelas2017}. The KHI creates a turbulent layer at the loop edges, where resonant absorption and phase mixing can effectively transfer energy to smaller scales. However, even if enough energy is provided to the system, its heating would still be mainly localised in the edge of the loop. 

Recently, decayless low-amplitude kink oscillations have been discovered in coronal loops \citep{nistico2013,anfinogentov2015}. The KHI could play an important role in this physical phenomenon, since the observations suggest that the decayless waves are also standing waves with an average amplitude of $\sim 0.2$ Mm, but lower than $1$ Mm . \citet{antolin2016} have proposed line-of-sight (LOS) effects due to the KHI and limits in the spatial resolution of our observations, as the cause of this observed decayless motion. Another proposal is the development of standing waves through a driving mechanism near the loop footpoints \citep{nakariakov2016}, like those simulated in \citet{karampelas2017}.

In the current study, we have expanded on our previous work, aiming to model the low-amplitude, decayless kink waves in active region coronal loops, driven by footpoint motions. In our previous study \citep{karampelas2017}, we had concentrated on the longitudinal dependence of the heating rate by the Transverse Wave Induced Kelvin-Helmholtz (TWIKH) rolls. Here, however, we have concentrated on an interesting result about the spatial evolution of a loop cross section for standing kink waves generated by footpoint drivers. In particular, we are going to study the effects of the TWIKH rolls on the cross-sectional density profile and the location of wave heating in the cross section of the loop.

\section{Numerical model}
We simulate footpoint-driven transverse waves in a $3D$ straight, density-enhanced magnetic flux tube, in a low-$\beta$ coronal environment. Our setup follows the work of \citet{karampelas2017}, with a loop length $L=200$ Mm and radius $R=1$ Mm. The loop density is equal to $\rho_i=2.509 \cdot 10^{-12}$ kg/m$^3$, and the loop temperature is $T_i = 9 \cdot 10^5$ K. The index i (e) denotes internal (external) values. The magnetic field is $B_z=22.8$ G and the plasma $\beta=0.018$. The density profile consists of the continuous function:
\begin{equation}
\rho(x,y) = \rho_e + 0.5\, \left( \rho_i - \rho_e \right) \lbrace 1-\tanh\left( \left( \dfrac{\sqrt{x^2+y^2}}{R}-1\right) \,b \right) \rbrace, 
\end{equation}
where $x$ and $y$ denote the co-ordinates in the plane perpendicular to the loop axis, $z$  along its axis. For $b=20$ the inhomogeneous layer has a width $\ell \approx 0.3 R$. The density ratio of $\rho_e/\rho_i = 1/3$, inspired by observational data in \citet{aschwanden2003VD}, leads to swift transfer of energy from transverse to azimuthal motions. In \citet{karampelas2017} it was found that the dynamics of the oscillations are not sensitive to the value of the temperature ratio. In the current setup we have effectively modelled a coronal loop during its cooling phase, by choosing a gradient of $T_i /T_e =1/3$. The external and internal Alfv\'{e}n speeds are equal to $\upsilon_{Ae}=2224$ km s$^{-1}$ and $\upsilon_{Ai}=1284$ km s$^{-1}$.

As in the previous study, we have used the MPI-AMRVAC code \citep{porth2014}, with an effective resolution of $512 \times 256 \times 64$. Since our domain dimensions are $(x,y,z) = (16,8,100)$ Mm, the cell dimensions are $31.25 \times 31.25 \times 1562.5$ km. The numerical resistivity is estimated to correspond to a Lundquist number of $S\geq 2.1 \cdot 10^4$.

As before, the tube is driven from the footpoint ($z=0$ Mm) with the kink period of $P\simeq 2L/c_k\simeq 254$ s \citep{edwin1983wave}. The driver velocity, at the bottom boundary, is uniform inside the loop and time varying as follows:
\begin{equation}
\lbrace v_x,v_y \rbrace=\lbrace v(t),0 \rbrace = \lbrace v_0 \cos\left(2\pi t / P\right),0 \rbrace ,
\end{equation}
where $v_0$ km/s is the peak velocity amplitude. We considered two different cases, one for a driver with a peak velocity of $v_0=2$ km/s, and one with $v_0=0.8$ km/s. Outside the loop, the velocity follows the relation:
\begin{equation}
\lbrace v_x,v_y \rbrace = v(t)R^2 \lbrace \dfrac{x^2-y^2}{(x^2+y^2)^2}, \, \dfrac{2xy}{(x^2+y^2)^2}\rbrace,
\end{equation}
with a smooth transition region between the two areas, matching the outer layer of the cylindrical tube. 
Following \citet{karampelas2017}, we also set the velocity component parallel to the $z$ axis ($v_z$) antisymmetric at the bottom boundary, to prevent mass flow through it. The driver sets the values for $v_x$ and $v_y$ at the bottom boundary, while all the other quantities obey a Neumann-type, zero-gradient condition there. Using the given driver frequency ensures that the superposing propagating waves (originating from each footpoint) form the fundamental standing kink mode for our tube. Taking advantage of the symmetric nature of the kink mode, we kept $v_z$, $B_x$, and $B_y$ antisymmetric in the $x-y$ plane at the top boundary (apex, $z=100$ Mm), while all the other quantities are symmetric. Furthermore, through the symmetric nature of our driver we set $v_y$ and $B_y$ antisymmetric in the $x-z$ plane, while the other quantities are symmetric. Therefore, we only simulated one fourth of our tube, as shown in Fig. \ref{fig:tube}. In the rest of the boundaries, a zero-gradient condition was used for all the quantities.

\section{Results}
For the rest of our analysis we focus on the sub-region of our computational domain, defined by $0 \leq z \leq 100$ Mm, $\vert x \vert \leq 2.33$ Mm and $y \leq 2.33$ Mm, where the resolution is the highest. We ran a simulation for a total time of ten driving periods ($10 \, P \sim 2540$ s).

\begin{figure}
\resizebox{\hsize}{!}{\includegraphics[trim={0cm 0cm 6cm 0cm},clip,scale=0.1]{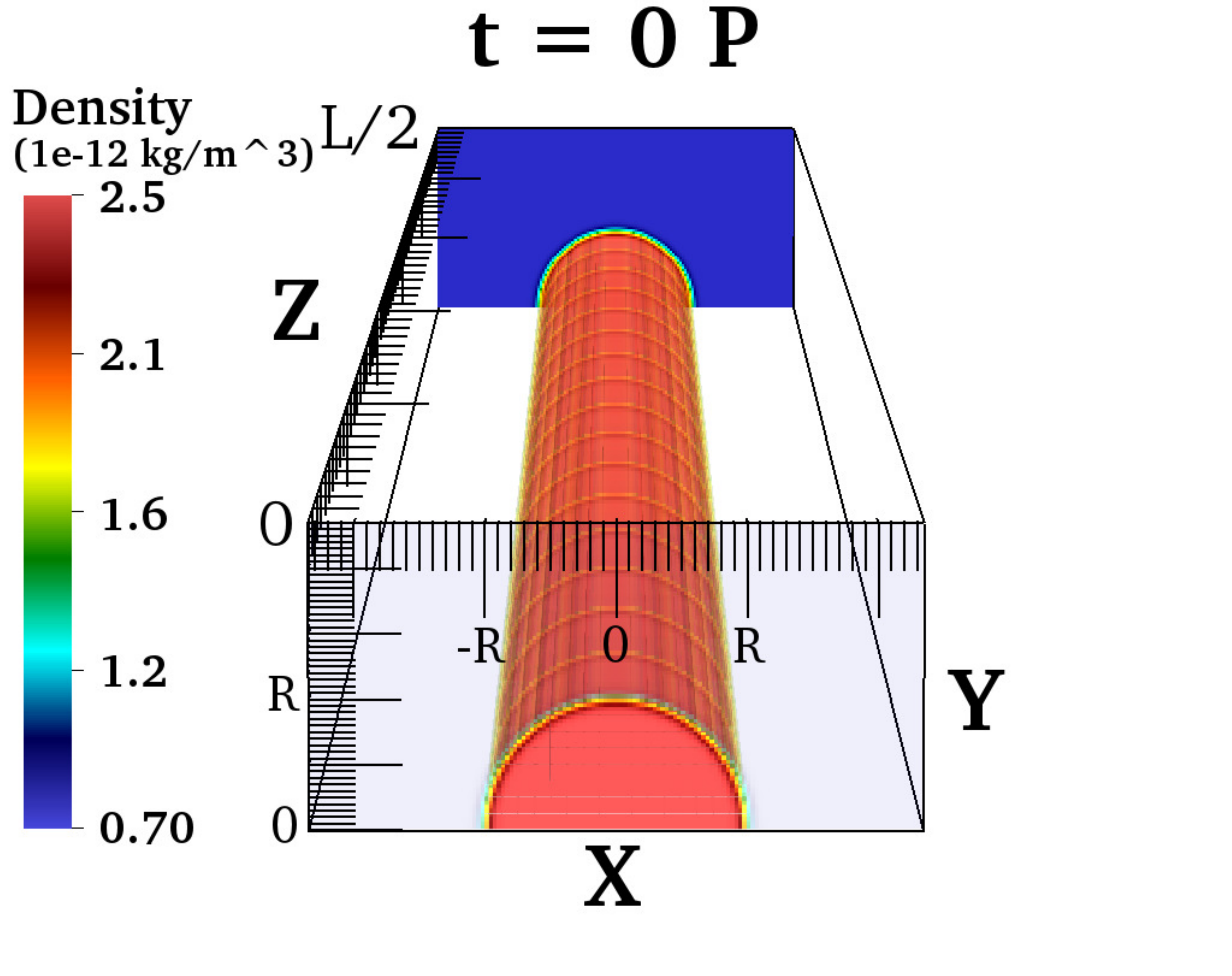}
\includegraphics[trim={7cm 0cm 6cm 0cm},clip,scale=0.1]{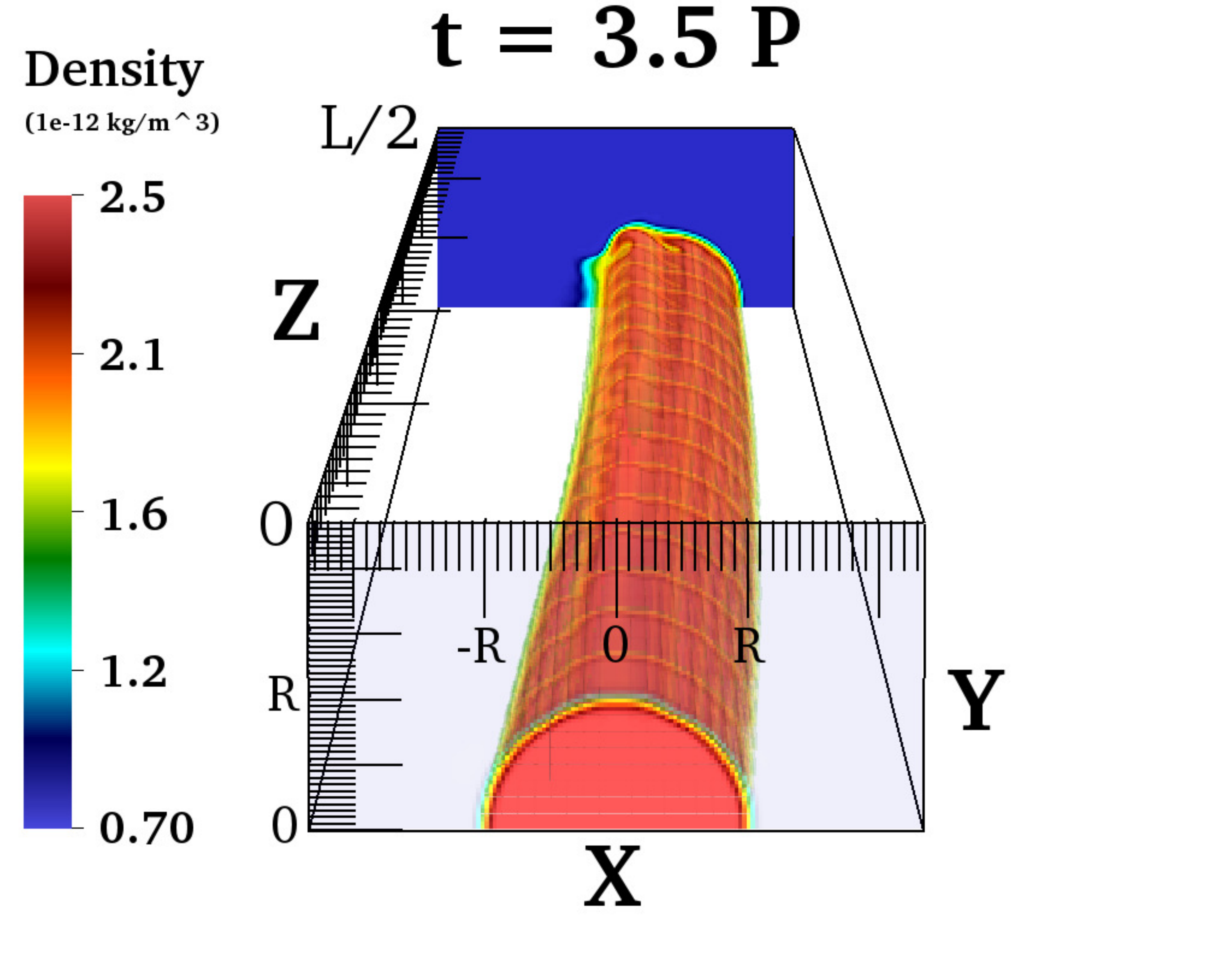}}
\resizebox{\hsize}{!}{\includegraphics[trim={0cm 0cm 6cm 0cm},clip,scale=0.1]{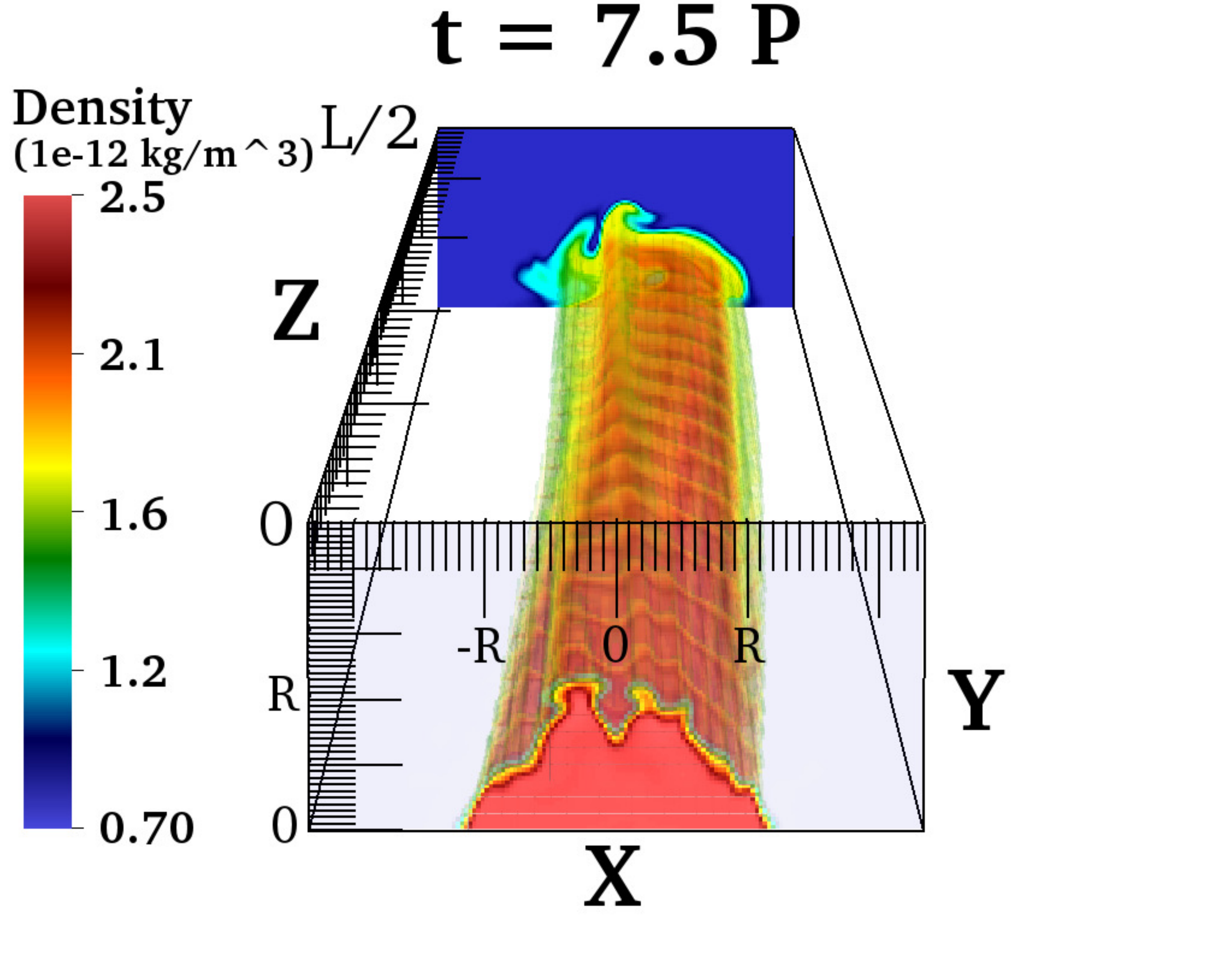}
\includegraphics[trim={7cm 0cm 6cm 0cm},clip,scale=0.1]{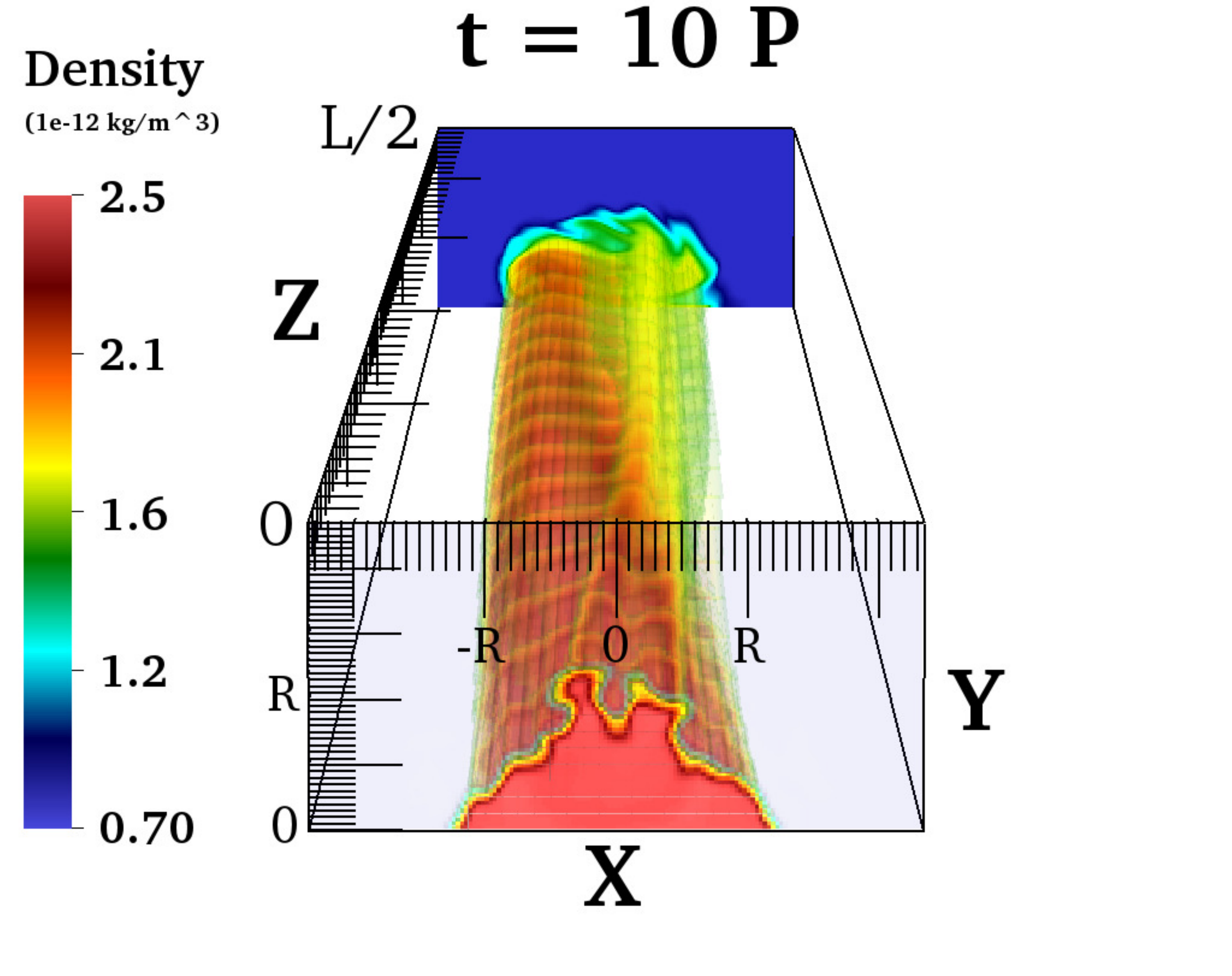}}
\caption{Density structure of the flux tube, for a driver with $v_0 = 2$ km/s. Snapshots are taken for times $t=0, \, 3.5 \, P, \, 7.5 \, P$ and $10 \, P $, where $P\simeq 254$ s is the driver period. The loop length is $L=200$ Mm and the loop radius is $R=1$ Mm. An animation of this figure, showing the oscillation for our model, is available online.}\label{fig:tube}
\end{figure}

\begin{figure}
\resizebox{\hsize}{!}{\includegraphics[trim={0.9cm 0cm 4.1cm 0.4cm},clip,scale=0.2]{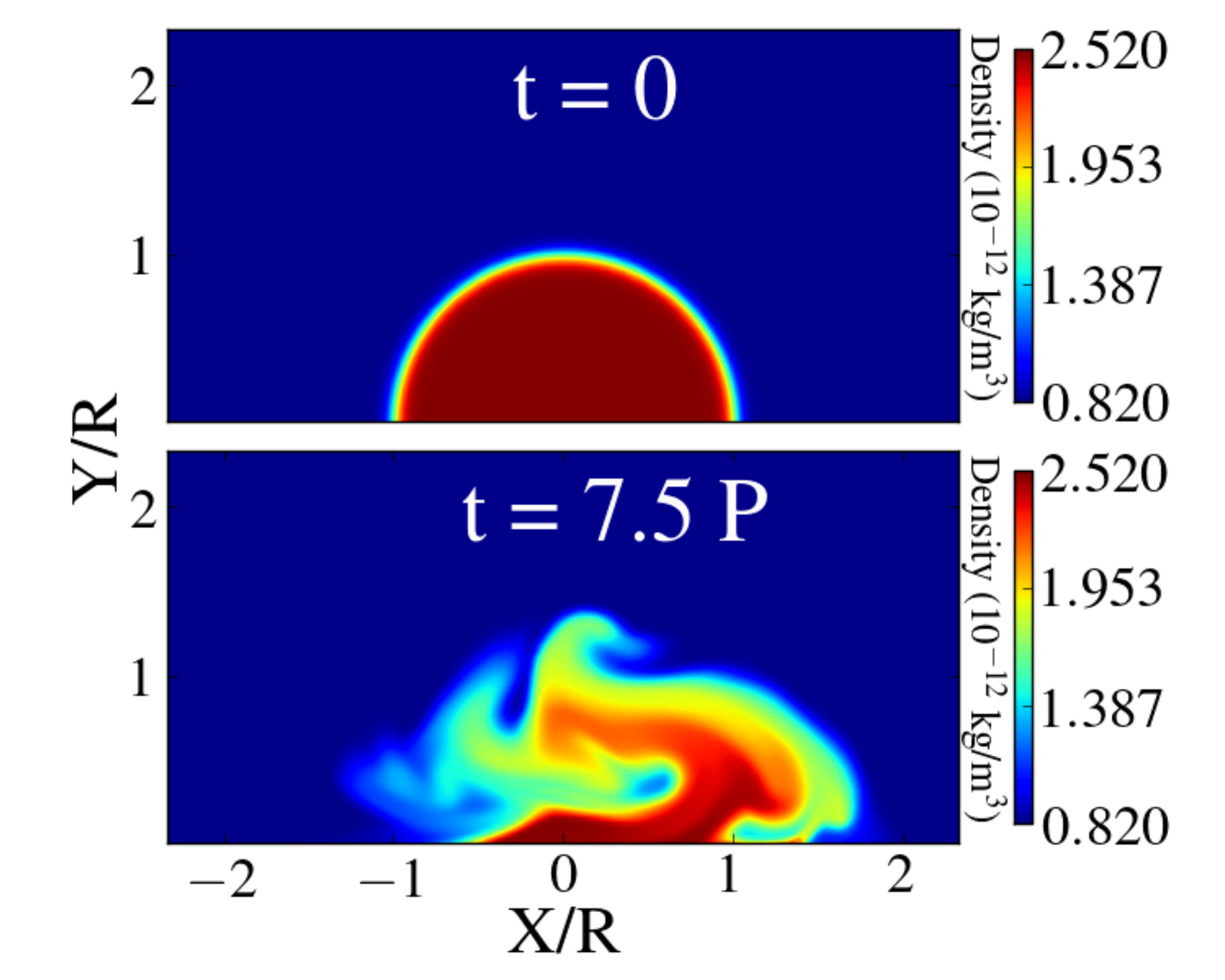}
\includegraphics[trim={2.5cm 0cm 0.8cm 0.4cm},clip,scale=0.2]{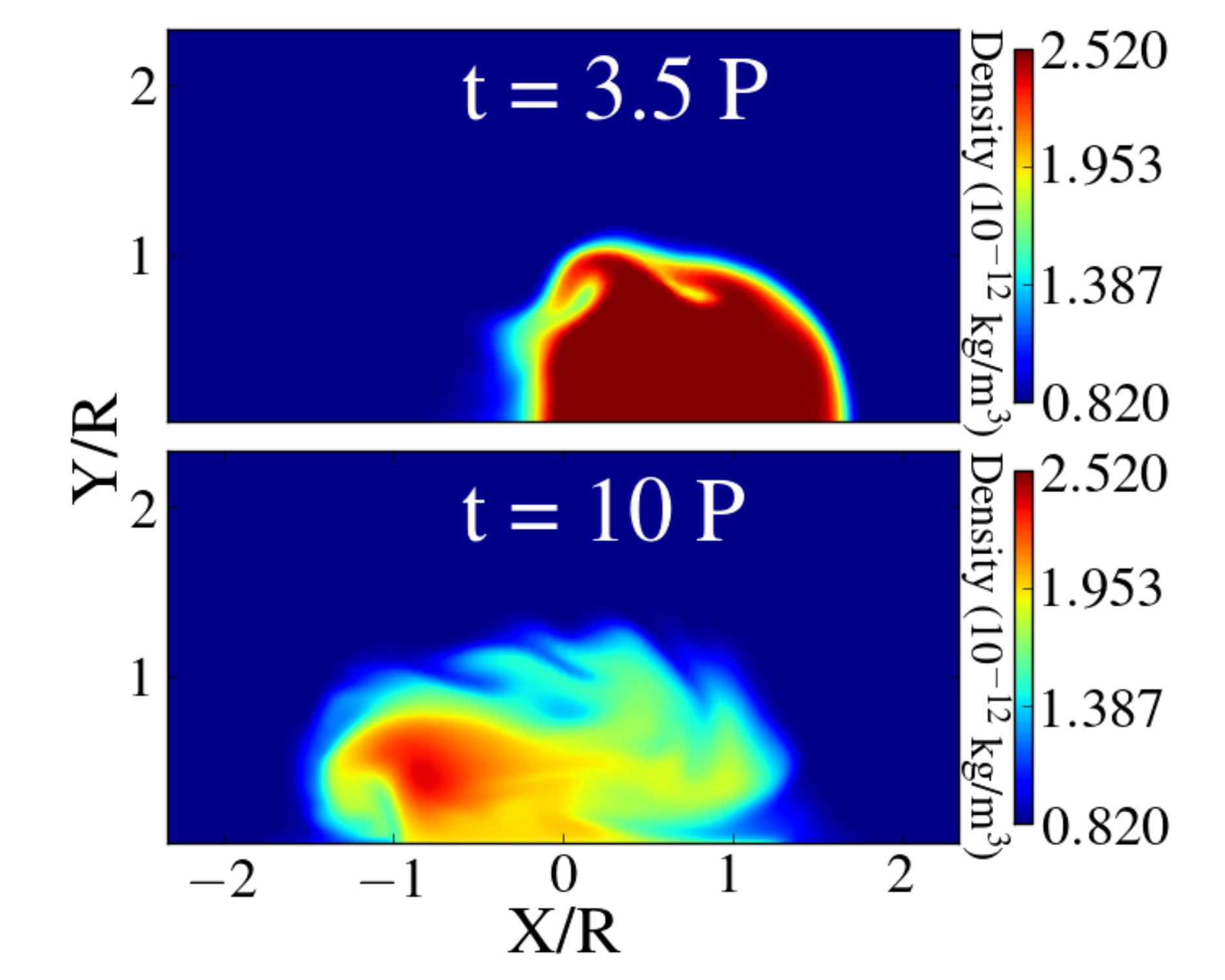}}
\caption{Snapshots of density ($10^{-12}$ kg/m$^3$) of the flux tube cross-section at the apex ($z=100$ Mm), for a driver with $v_0 = 2$ km/s. $P\simeq 254$ s is the driver period.}\label{fig:dens}
\end{figure}

\begin{figure}
\resizebox{\hsize}{!}{\includegraphics[trim={0.9cm 0cm 4.1cm 0.4cm},clip,scale=0.2]{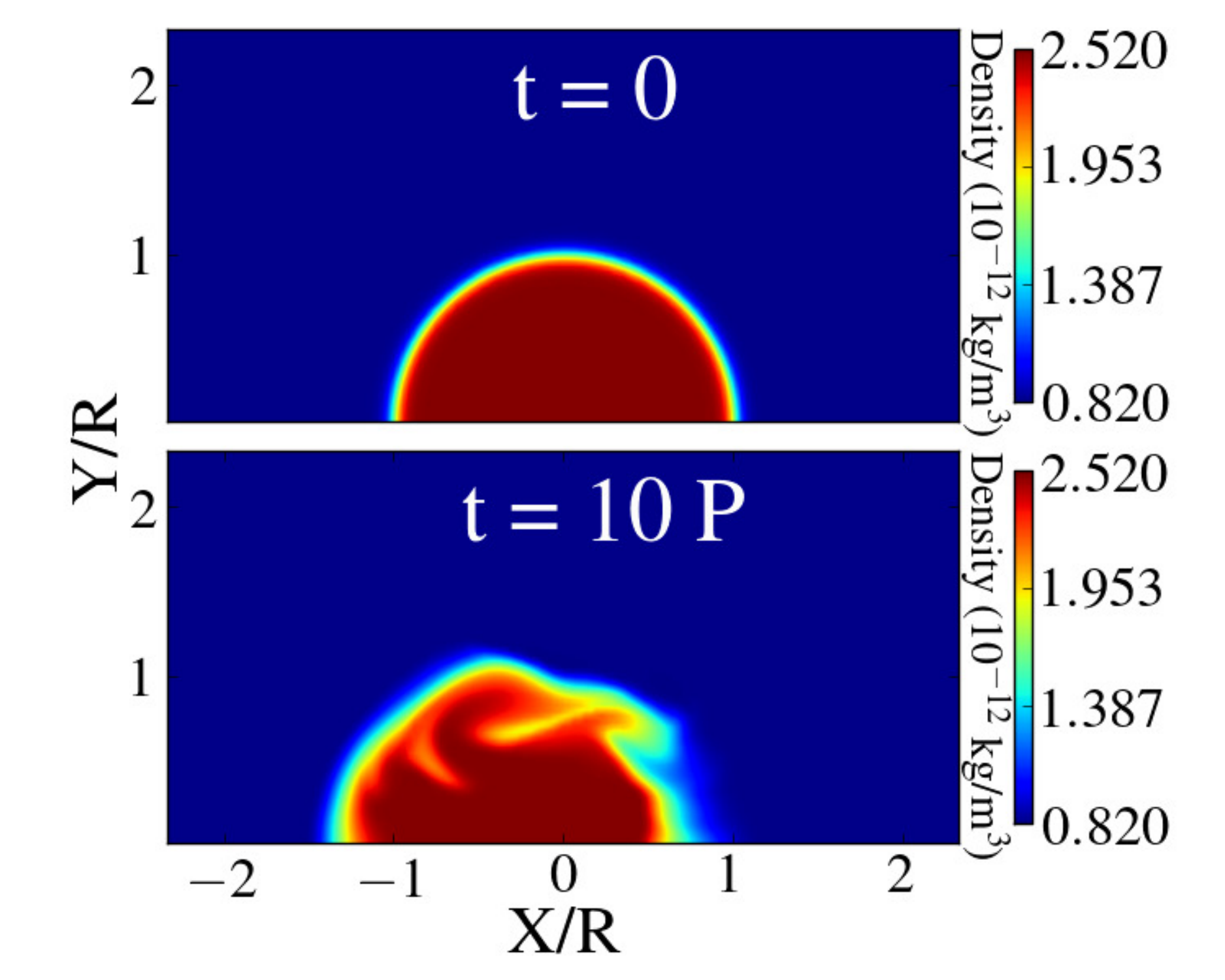}
\includegraphics[trim={2.5cm 0cm 0.8cm 0.4cm},clip,scale=0.2]{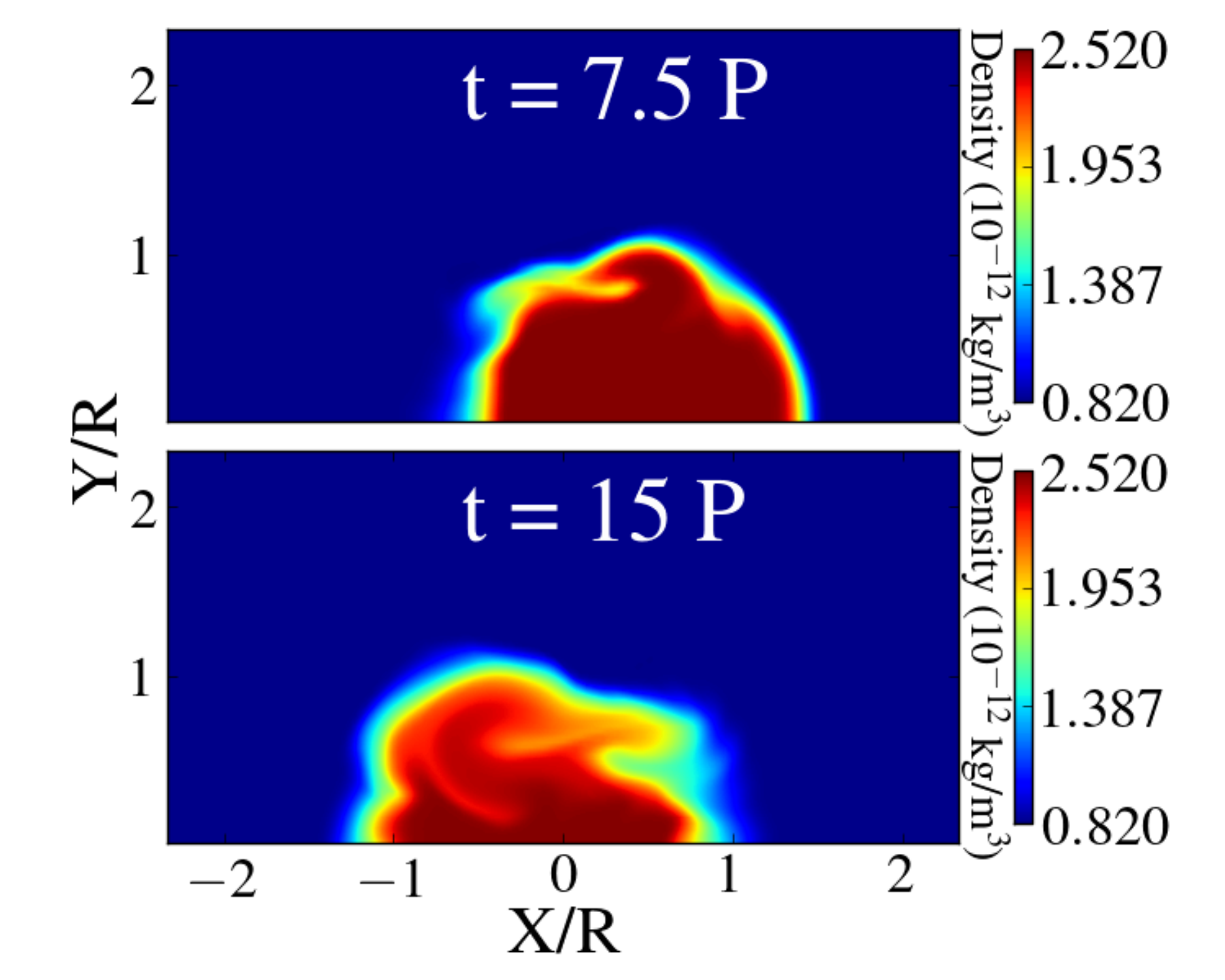}}
\caption{Same as Fig. \ref{fig:dens}, but for a driver with $v_0 = 0.8$ km/s.}\label{fig:densslow}
\end{figure}

The loop apex, which is the location of the antinode of the $x$-velocity, is Kelvin-Helmholtz unstable, as expected from theory \citep{heyvaerts1983,zaqarashvili2015ApJ}. As we see in Fig. \ref{fig:dens}, the KHI manifests at the apex. In addition to that, we have the formation of spatially extended eddies, the TWIKH rolls. A similar feature can be seen in Fig. \ref{fig:densslow}, where we show the density cross-section of our tube at the apex for the driver with peak velocity $v_0=0.8$ km/s. The TWIKH rolls are now less prominent, due to the smaller driver velocity amplitude, which leads to slower increase of the shear velocities at the apex. However, letting the driver act for more periods also leads to a deformed cross section at the apex.

\begin{figure}
\centering
\resizebox{\hsize}{!}{\includegraphics[trim={0cm 4.2cm 3.5cm 4.1cm},clip,scale=0.28]{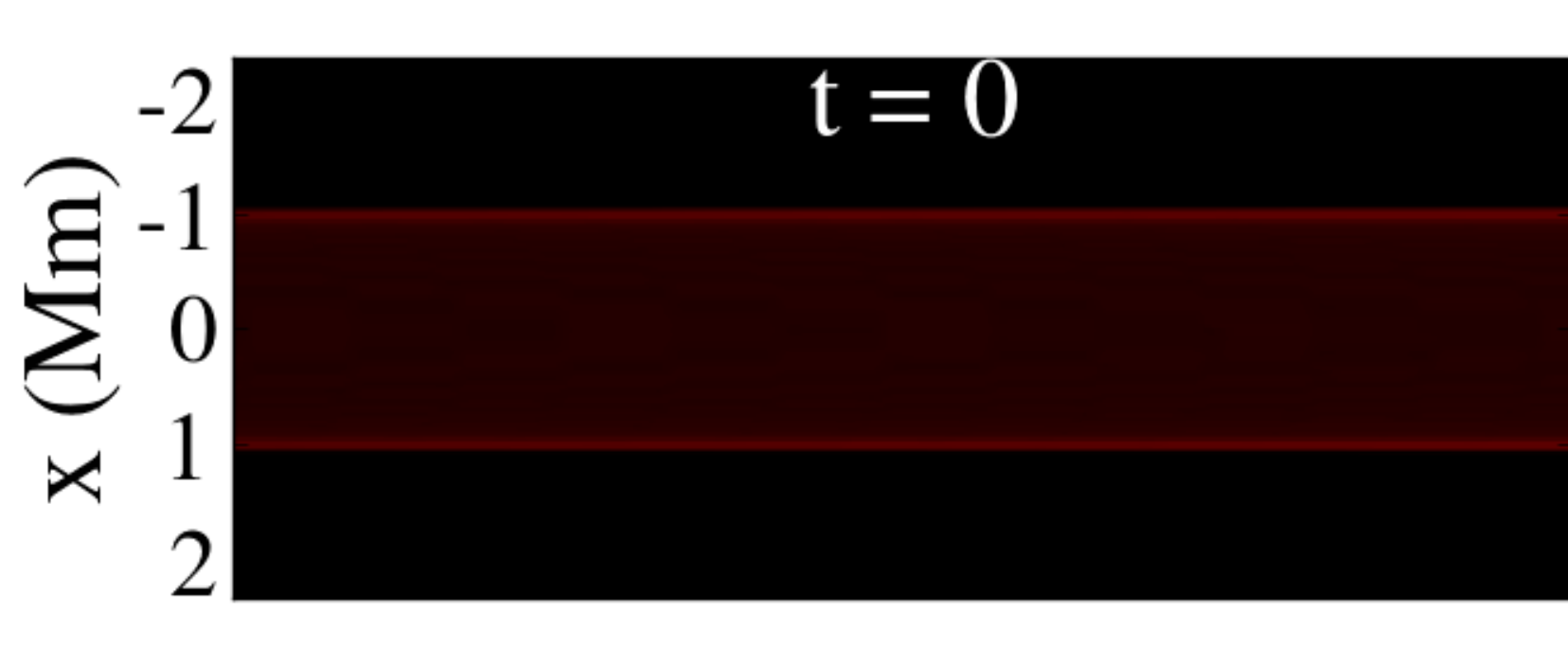}
\includegraphics[trim={2.2cm 4.2cm 0cm 4.1cm},clip,scale=0.28]{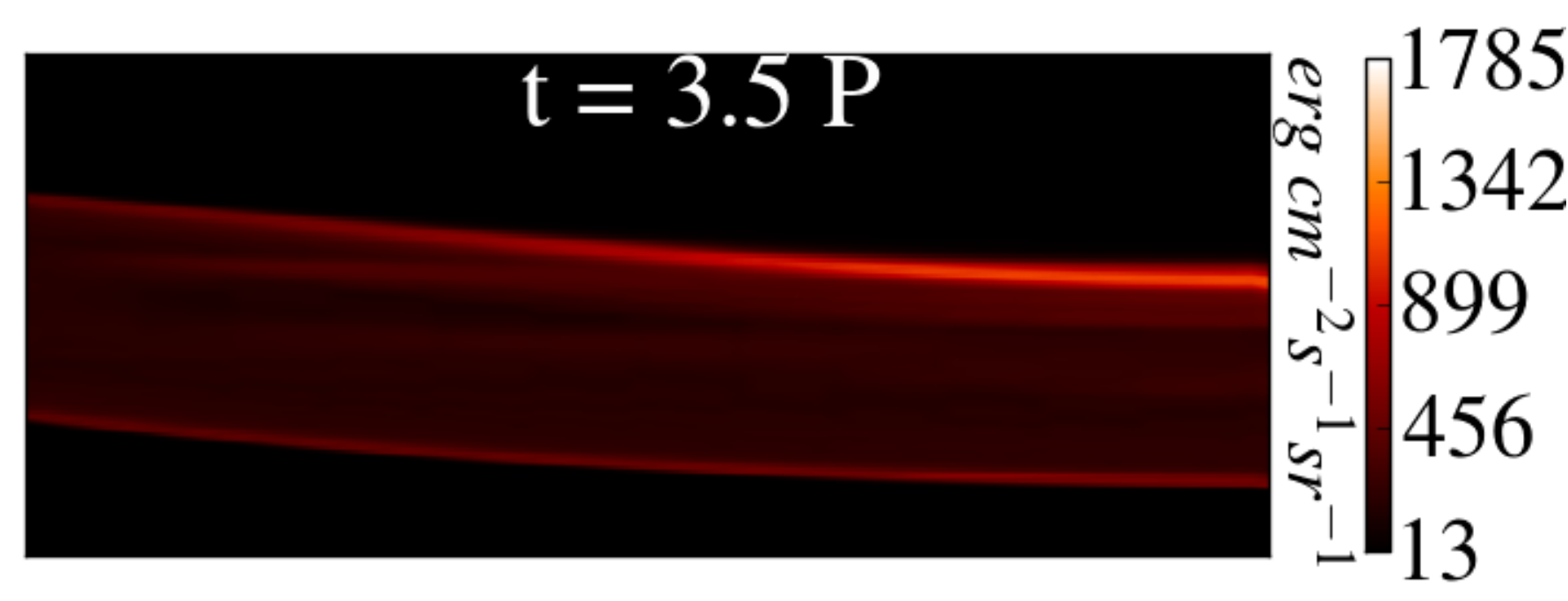}}
\resizebox{\hsize}{!}{\includegraphics[trim={0cm 2.5cm 3.5cm 4.3cm},clip,scale=0.28]{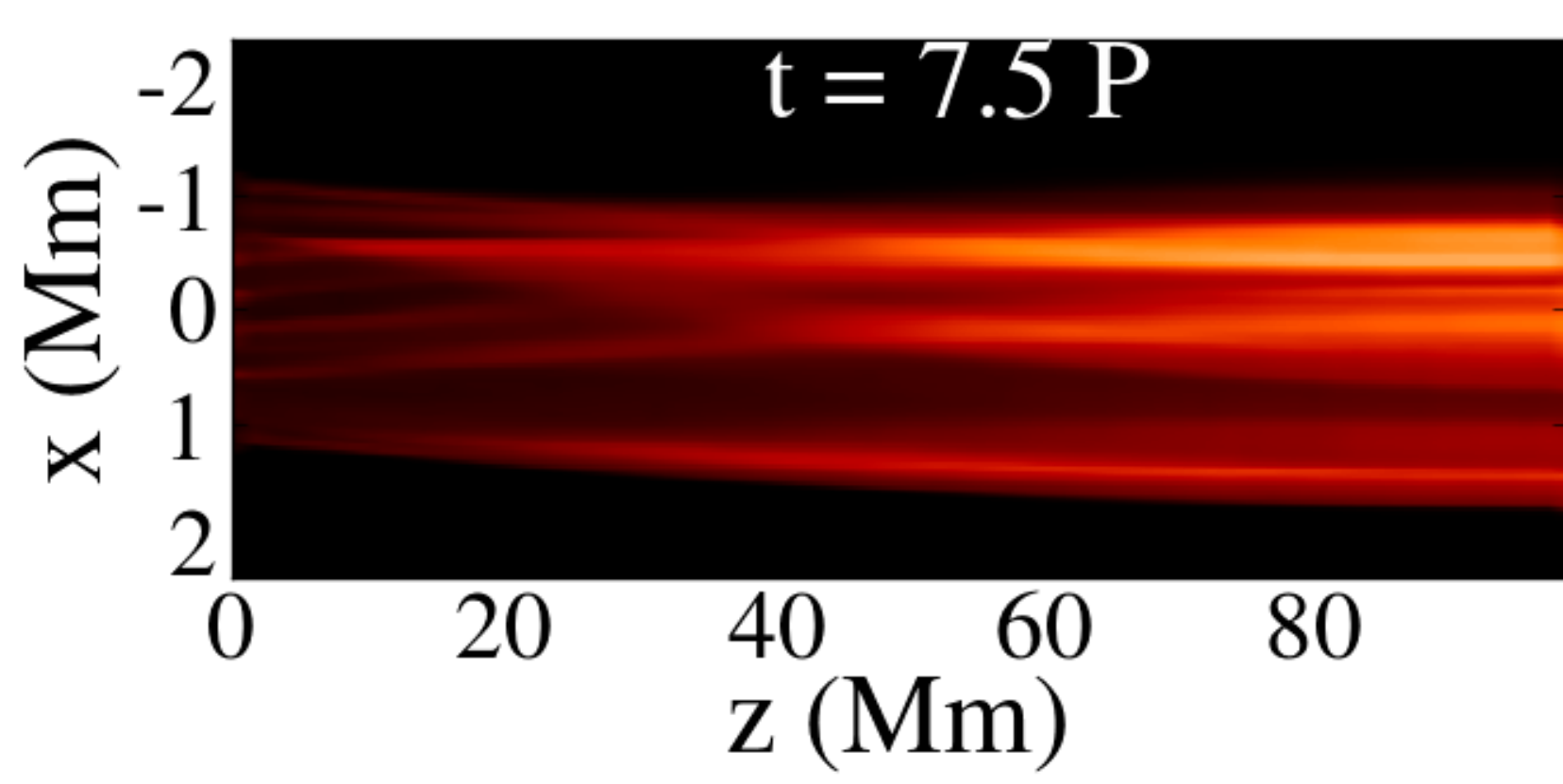}
\includegraphics[trim={2.2cm 2.5cm 0cm 4.3cm},clip,scale=0.28]{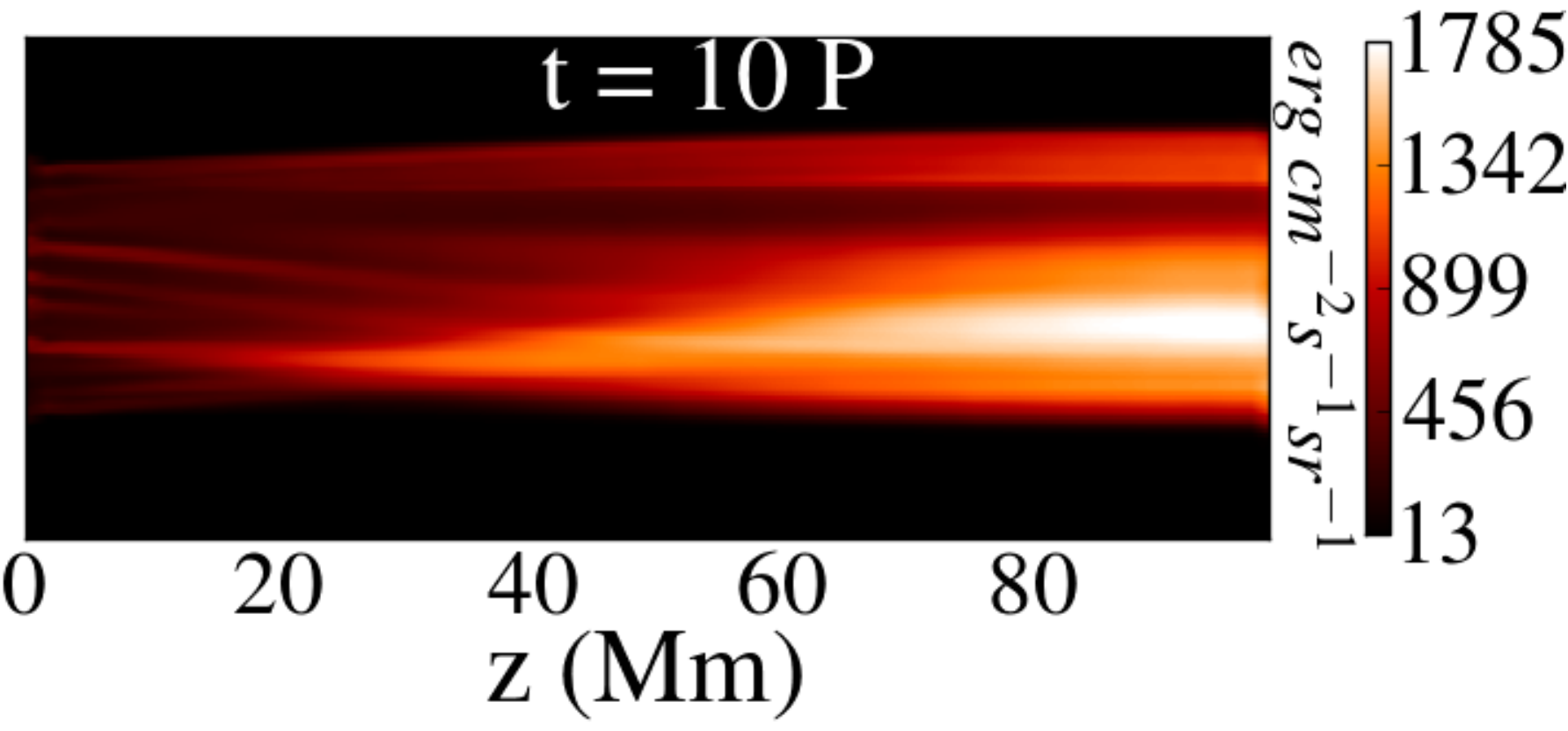}}
\caption{Forward modelling images of the integrated emission intensity (in erg cm$^{-2}$s$^{-1}$sr$^{-1}$) of the tube for the $193.509 \, \AA$ line. The observer is at a $0^\circ$ LOS angle, perpendicular to the oscillatory motion. Half the loop length is modelled ($z=0-100$ Mm). The driver peak velocity is $v_0=2$ km/s, and $P\simeq 254$ s is the driver period.}\label{fomotube}
\end{figure}

By performing forward modelling with the FoMo code \citep{fomo2016}, we observe the manifestation of strand-like structures due to the out-of-phase movements of the TWIKH rolls, which we also see in Fig. \ref{fig:tube}. These strands, resembling those in \citet{antolin2016} for the impulsively excited standing kink waves, are depicted in Fig. \ref{fomotube}. Here, we present snapshots of the emission intensity for the Fe XII $193.509\, \AA$ line, at times $t=0, \, 3.5 \, P, \, 7.5 \, P$ and  $10 \, P $. In our setup, the chosen line is better suited to detect the hotter plasma at the loop edge \citep{antolin2017}. The colour-scale is limited between the minimum and the maximum intensity values of the integrated intensity of the simulations. We consider a line-of-sight plane perpendicular to the loop axis and we set the LOS angle perpendicular to the oscillation direction equal to $0^\circ$. By choosing the given LOS angle and studying only the emission intensity, we avoid any missed emission from performing forward modelling in only half the loop cross-section.

\begin{figure*}
\centering
\resizebox{\hsize}{!}{\includegraphics[trim={0cm 2.2cm 0.5cm 0cm},clip,scale=0.25]{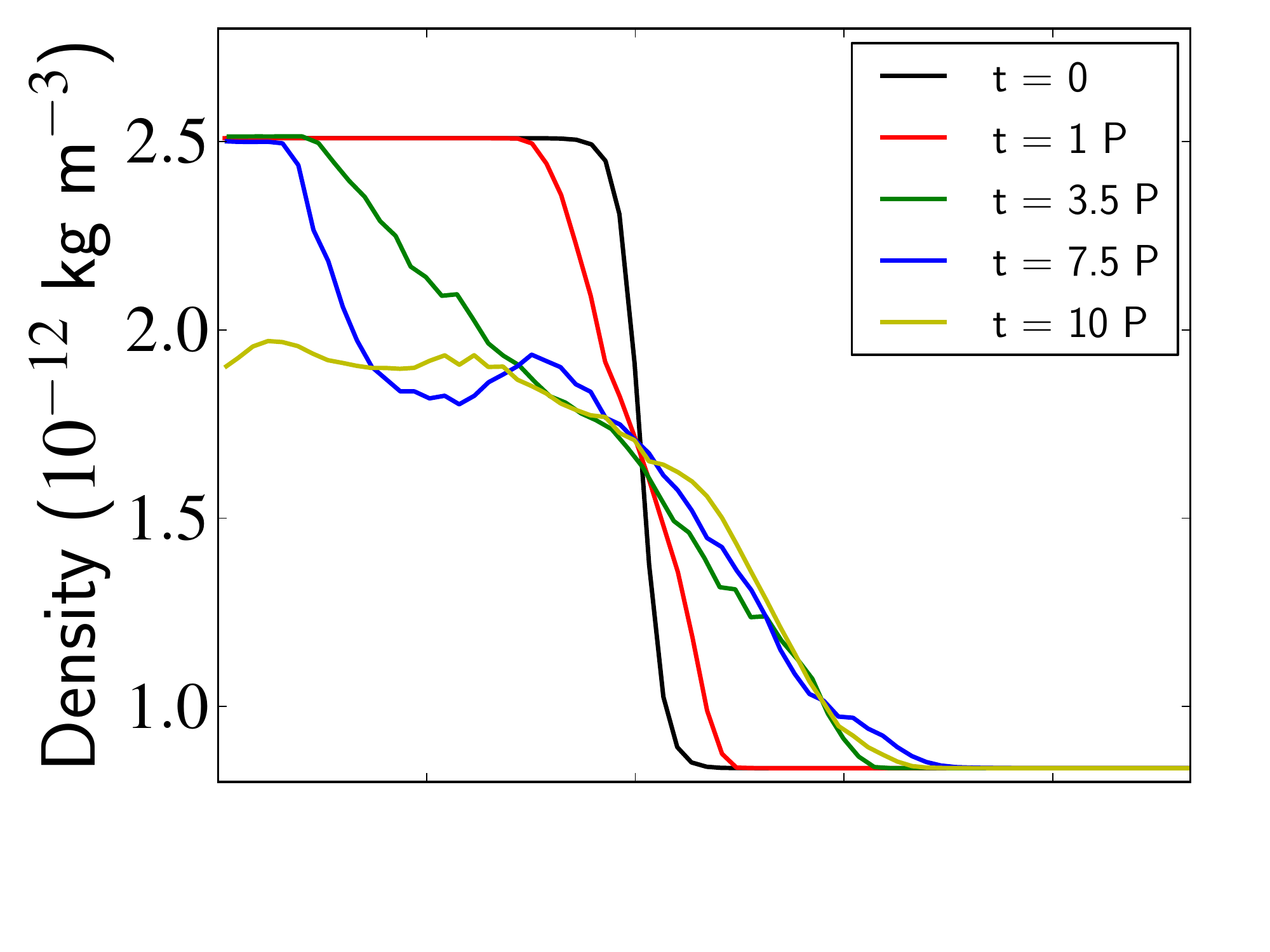}
\includegraphics[trim={0cm 2.2cm 0.5cm 0cm},clip,scale=0.25]{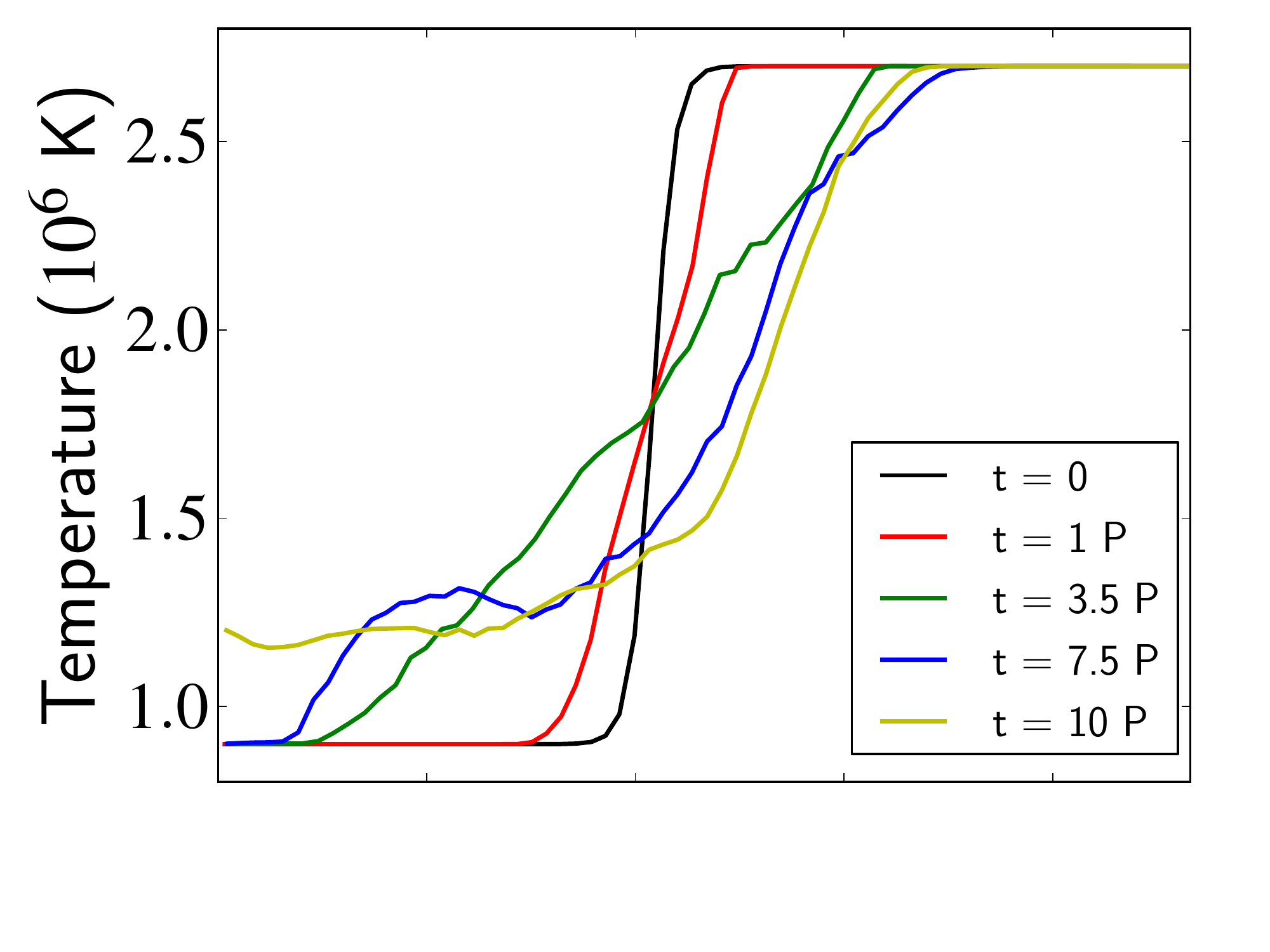}
\includegraphics[trim={0cm 2.2cm 0.5cm 0cm},clip,scale=0.25]{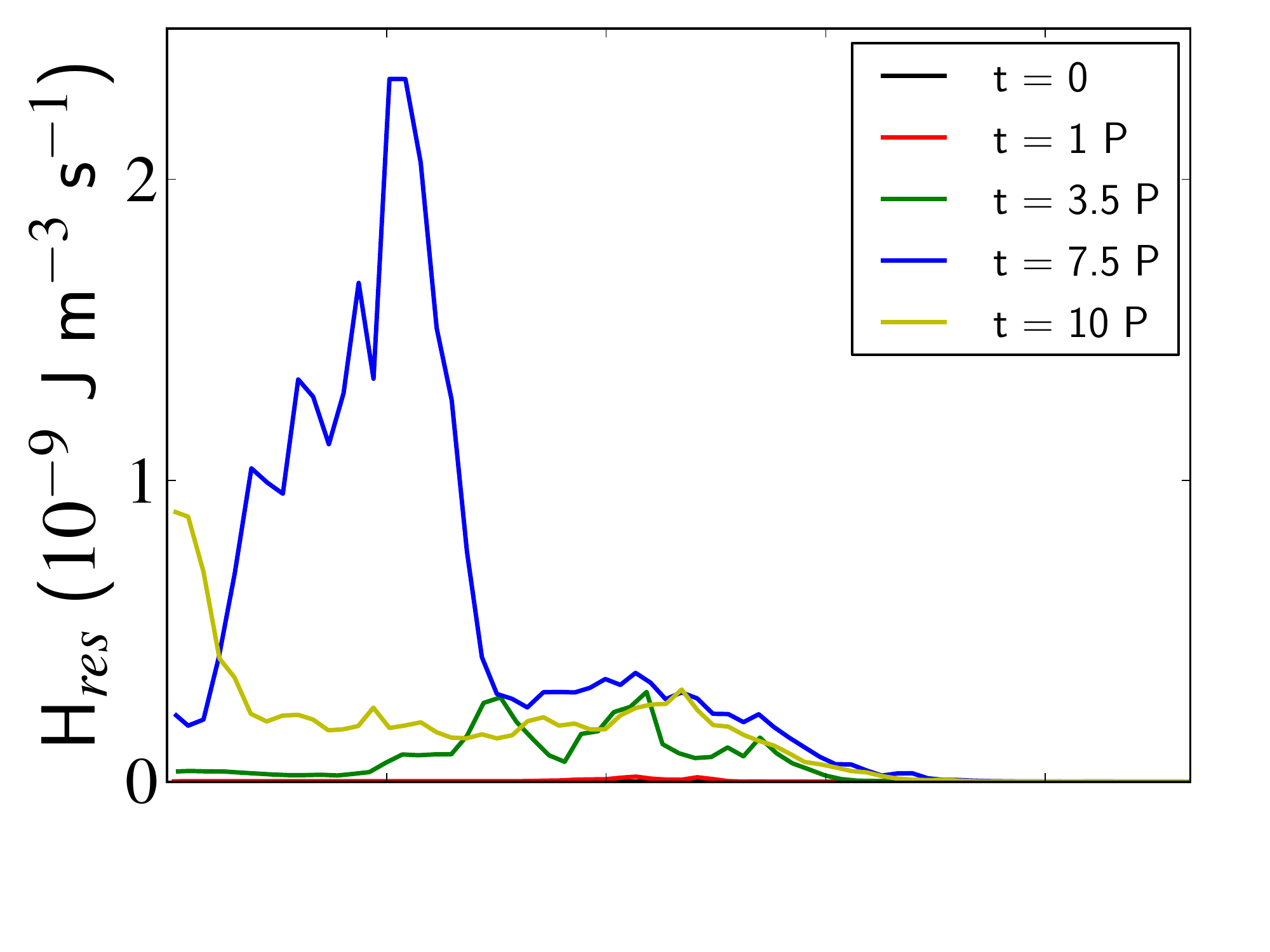}
\includegraphics[trim={0cm 2.2cm 0.5cm 0cm},clip,scale=0.25]{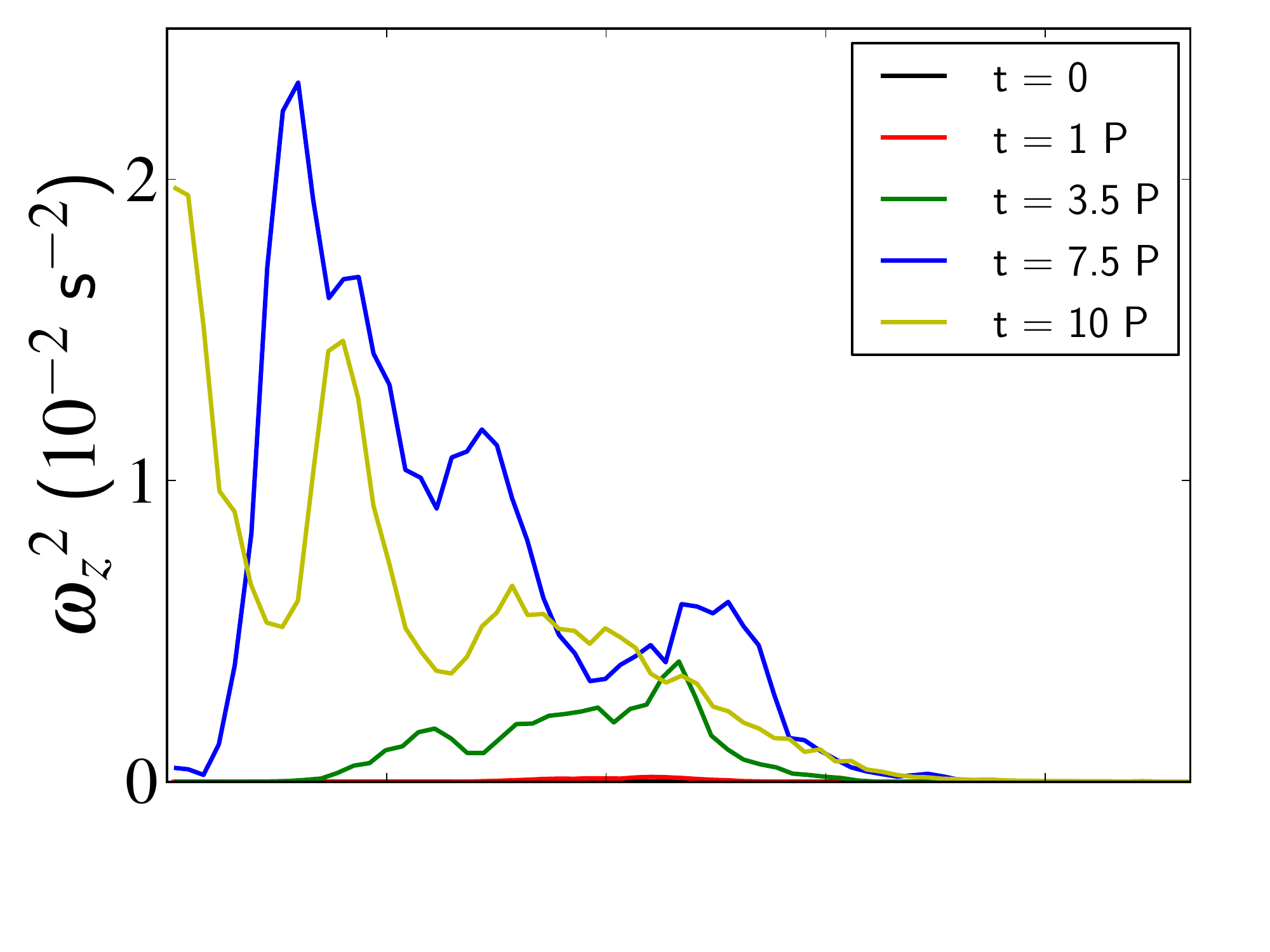}}
\resizebox{\hsize}{!}{\includegraphics[trim={0cm 0cm 0.5cm 0cm},clip,scale=0.25]{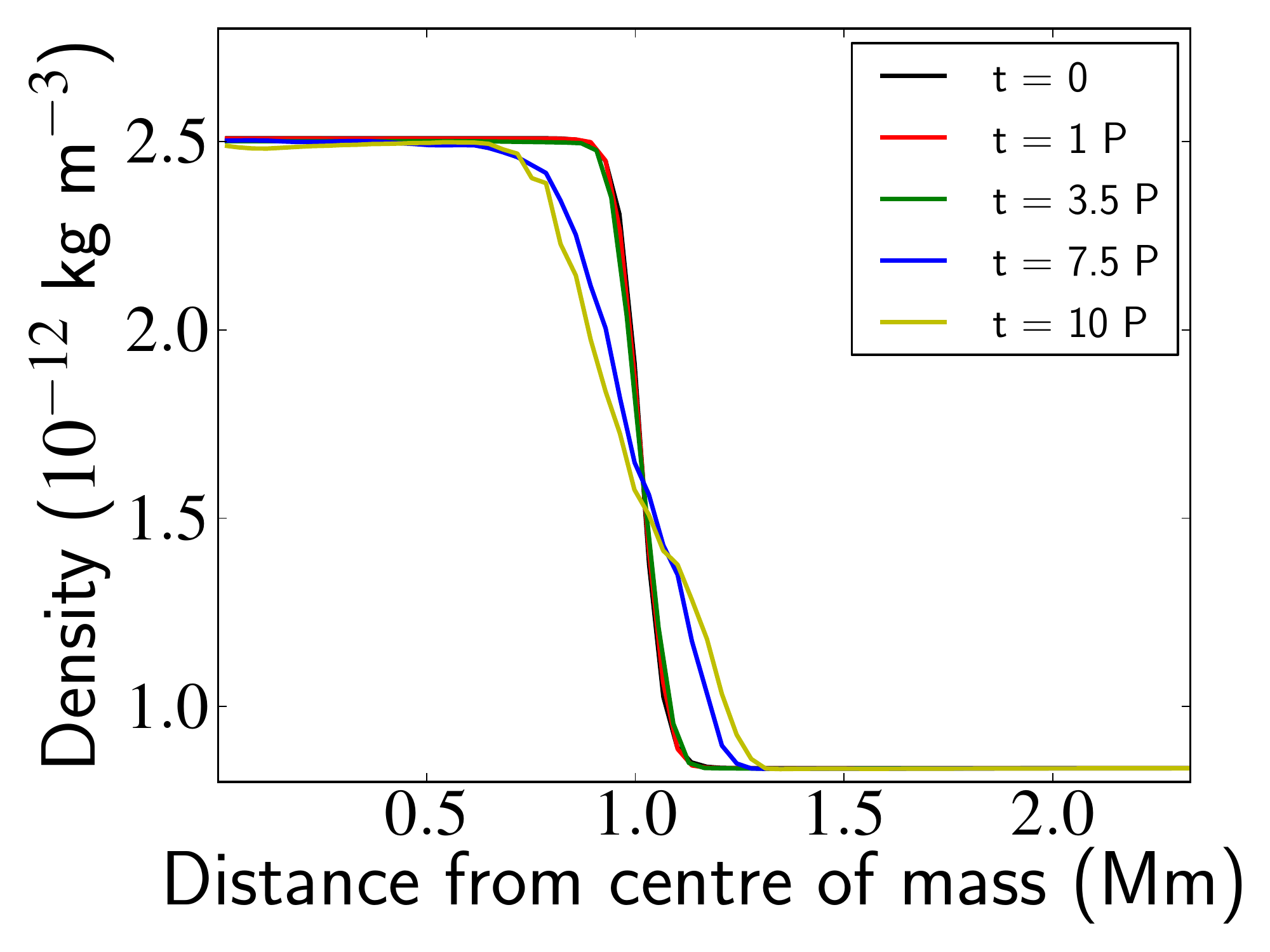}
\includegraphics[trim={0cm 0cm 0.5cm 0cm},clip,scale=0.25]{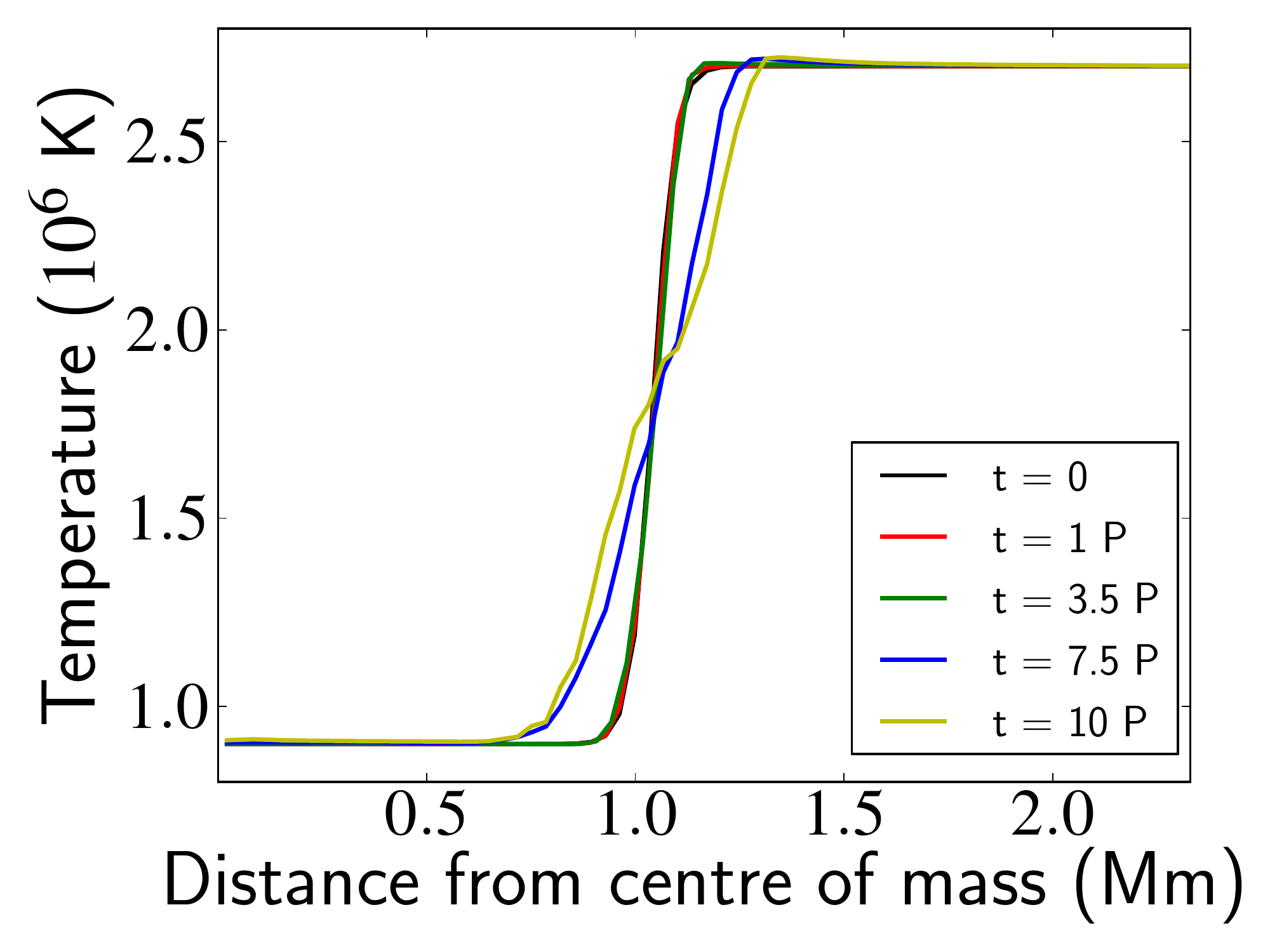}
\includegraphics[trim={0cm 0cm 0.5cm 0cm},clip,scale=0.25]{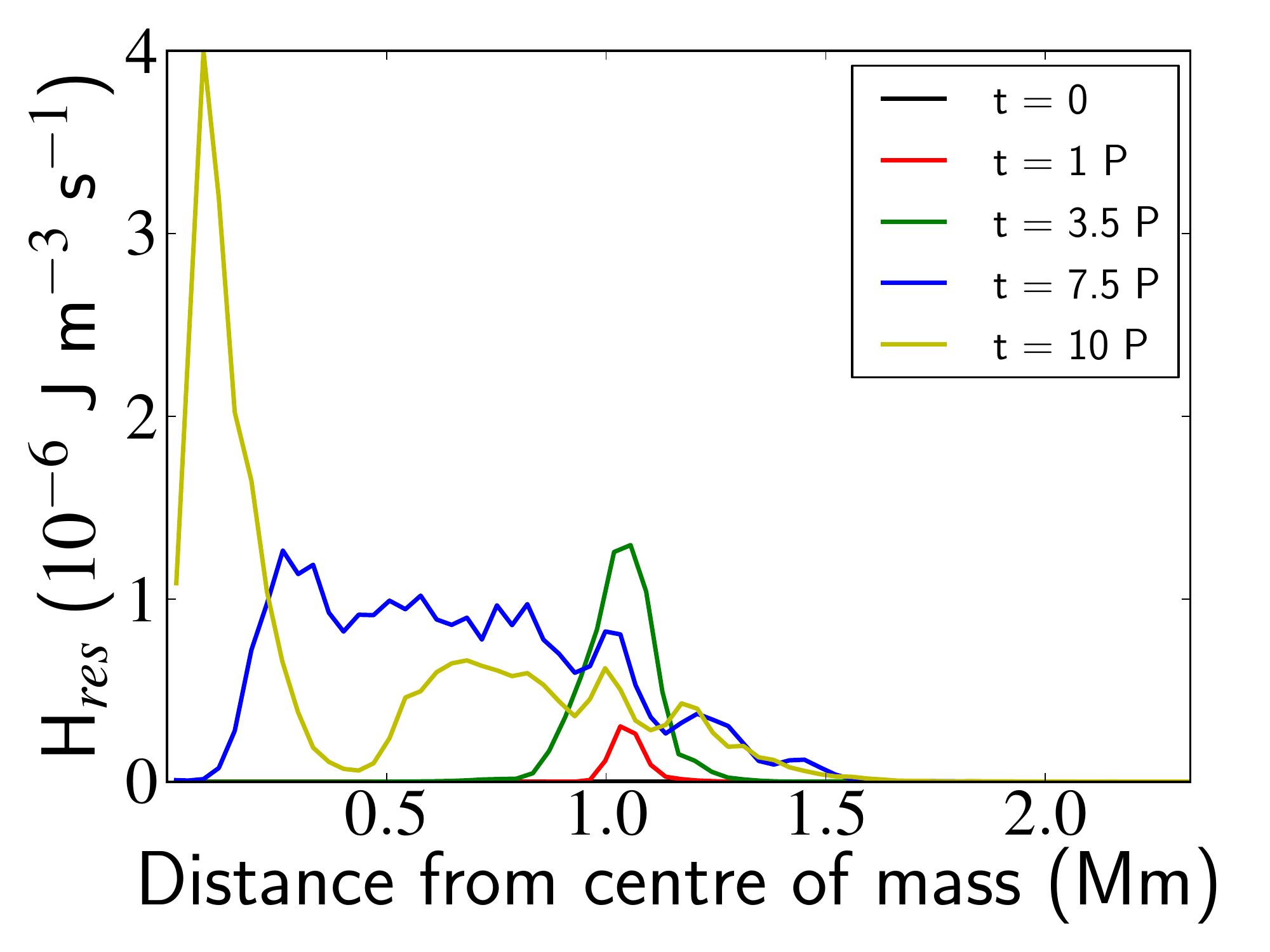}
\includegraphics[trim={0cm 0cm 0.5cm 0cm},clip,scale=0.25]{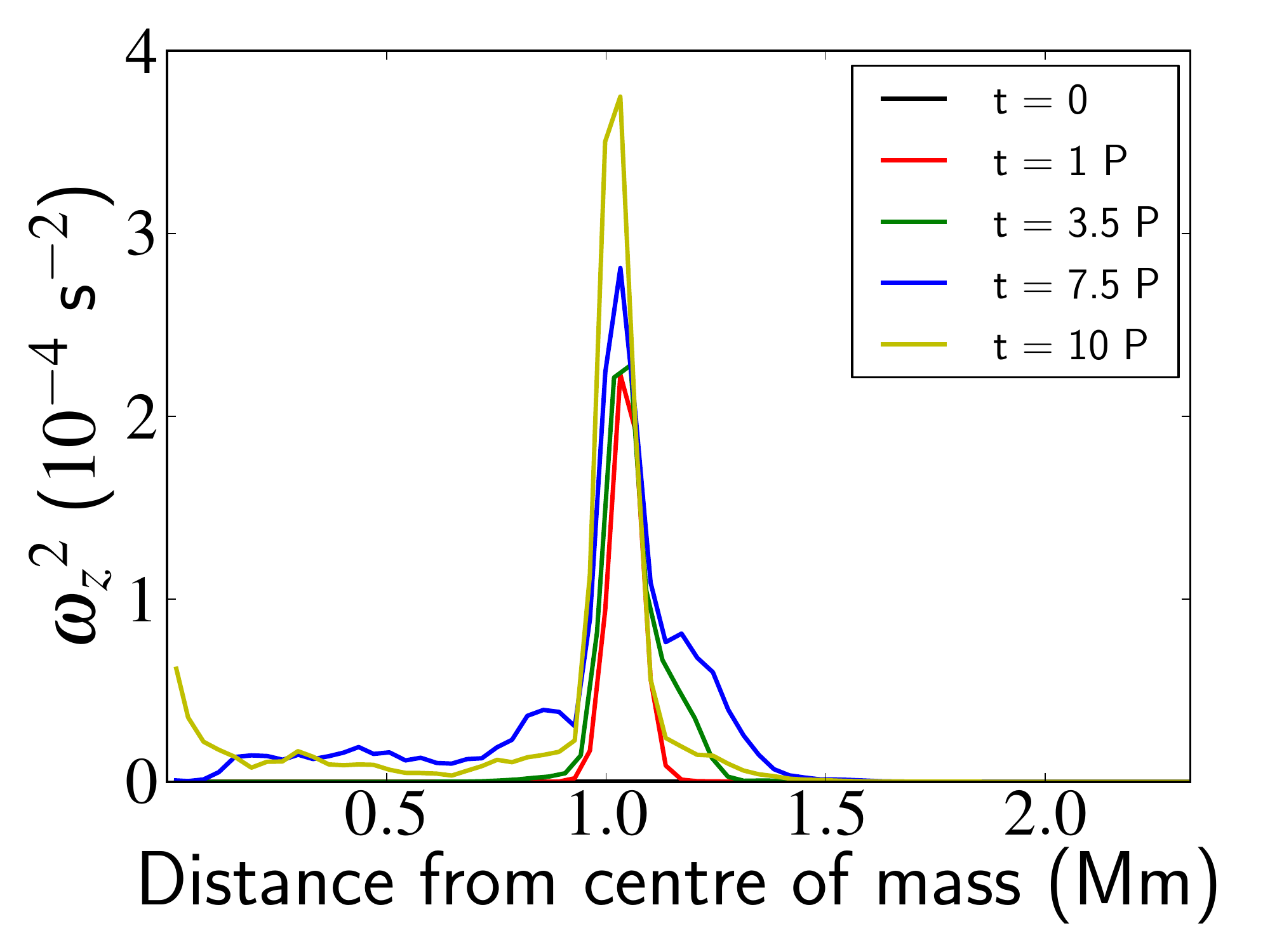}}
\caption{Profiles of the average density, temperature, resistive heating rate $H_{res}$ and vorticity $\omega_z^2$ as a function of the distance from the centre of mass. The driver peak velocity is $v_0 = 2$ km/s, and $P\simeq 254$ s is the driver period. The top row shows the profiles at the apex ($z=100$ Mm). The bottom row shows the profiles near the footpoint ($z=1$ Mm).}\label{fig:struct}
\end{figure*}

In Fig. \ref{fig:struct}, we plot the average density, temperature, resistive heating rate ($H_{res}$) and the heating by shear viscosity (traced here by square $z$-vorticity, $\omega_z^2$), as functions of distance from the centre of mass, both at the apex and near the footpoint. Initially, the values of $H_{res}$ and $\omega_z^2$ are zero, and remain small for the first few periods. At the apex, both the density and temperature spread across the cross section, as a result of the extended TWIKH rolls, effectively widening the loop boundary layer. Near the footpoint, suppression of the KHI results in less mixing, which is evident from the corresponding density and temperature profiles. The resistive heating rate peaks near the footpoint, in agreement with \citet{karampelas2017}. On the other hand, the viscous heating rate peaks near the loop. Radially, both of them are initially confined to the resonant layer, but later spread out over the entire loop cross section, as we can see in Fig. \ref{fig:vortHJ} for the $\omega_z^2$ at the apex and the $H_{res}$ at the footpoint, for snapshots at $t=7.5 \, P$.

\section{Discussion and conclusions}

Looking at the loop cross section at the apex (Fig. \ref{fig:dens}), it is interesting to see the gradual deformation of the loop, as the simulation reaches its final stages. In the current model, the site of the highest deformation is located in the area near the loop apex, because the $v_x, \, v_y$ velocity antinode and $B_x, \, B_y$ magnetic field node appear there. Previous works have shown that TWIKH rolls create a wide turbulent layer both for impulsive \citep{magyar2016damping} and driven \citep{karampelas2017} standing modes. Here our previous simulations were performed for an extended duration ($t_{max}=10$ P). The extended TWIKH rolls result in a completely deformed loop cross section, where extensive mixing is taking place across the entire loop. Thus the loop cross section at the apex becomes fully deformed. Studying the results in Fig. \ref{fig:densslow} (and as is intuitively clear), we see that the driver amplitude plays an important role in the development of the KHI and the evolution of the loop cross section. However, we find that the loop cross section also evolves to a fully deformed state, even for the very small driver amplitude ($v_0=0.8$ km/s). Therefore, it is probably safe to assume that coronal loops have a deformed cross section, if they are driven by transverse footpoint motions for a sufficiently long time.

The turbulent nature of the cross section has a profound impact on the radial structure of the loop. As we see from the top panels of Fig. \ref{fig:struct}, the density and temperature profiles near the apex seem to get smoothed over time, in these angle-averaged radial density and temperature profiles. This effect is most prominent between $50$ Mm and $100$ Mm (apex). Near the loop footpoint, however, the smearing of density and temperature profiles is less, because the TWIKH rolls are absent there.

\begin{figure}
\centering
\resizebox{\hsize}{!}{\includegraphics[trim={0cm 0.6cm 1.5cm 1.6cm},clip,scale=0.5]{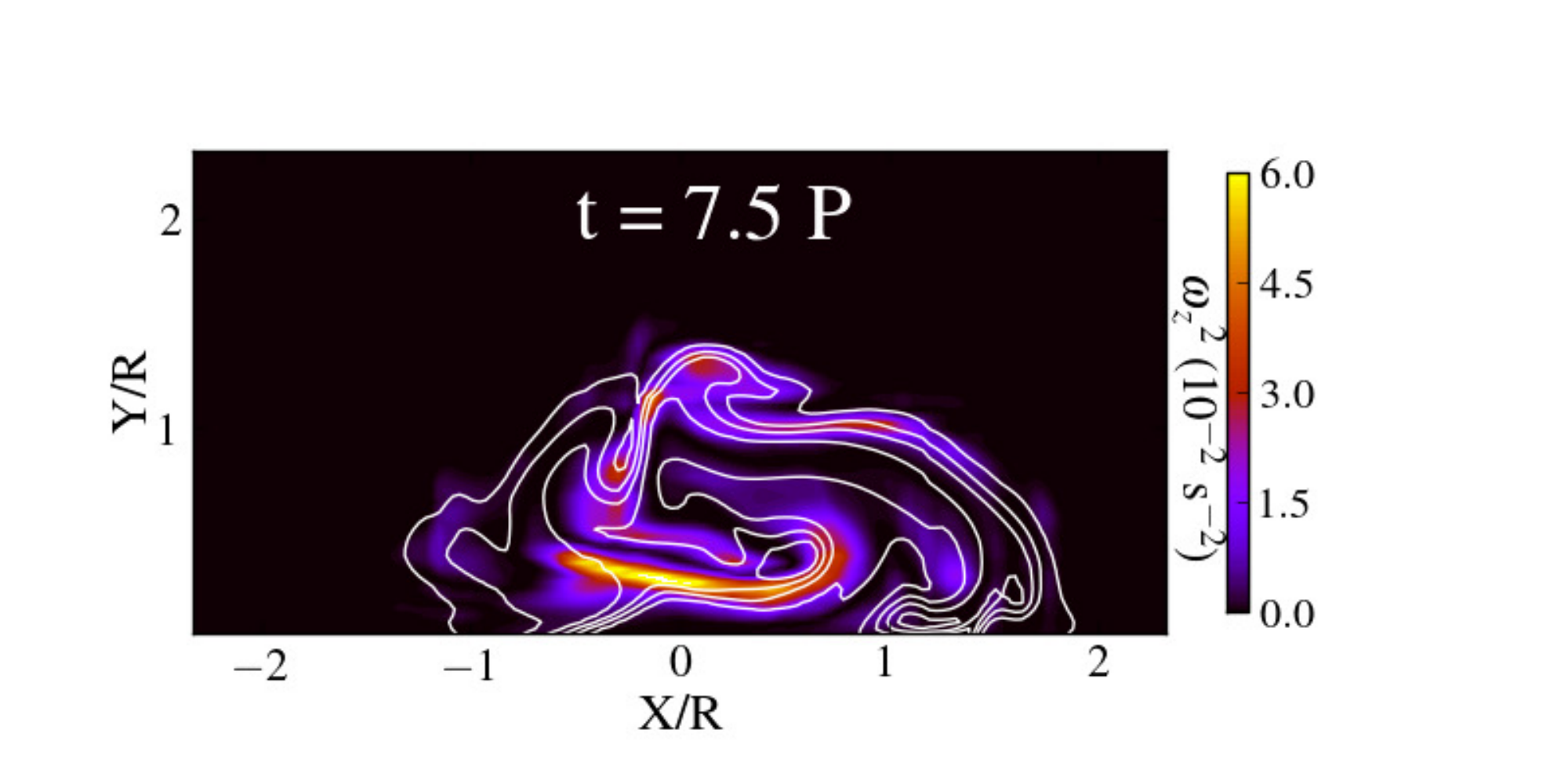}}
\resizebox{\hsize}{!}{\includegraphics[trim={0cm 0.6cm 1.5cm 1.6cm},clip,scale=0.5]{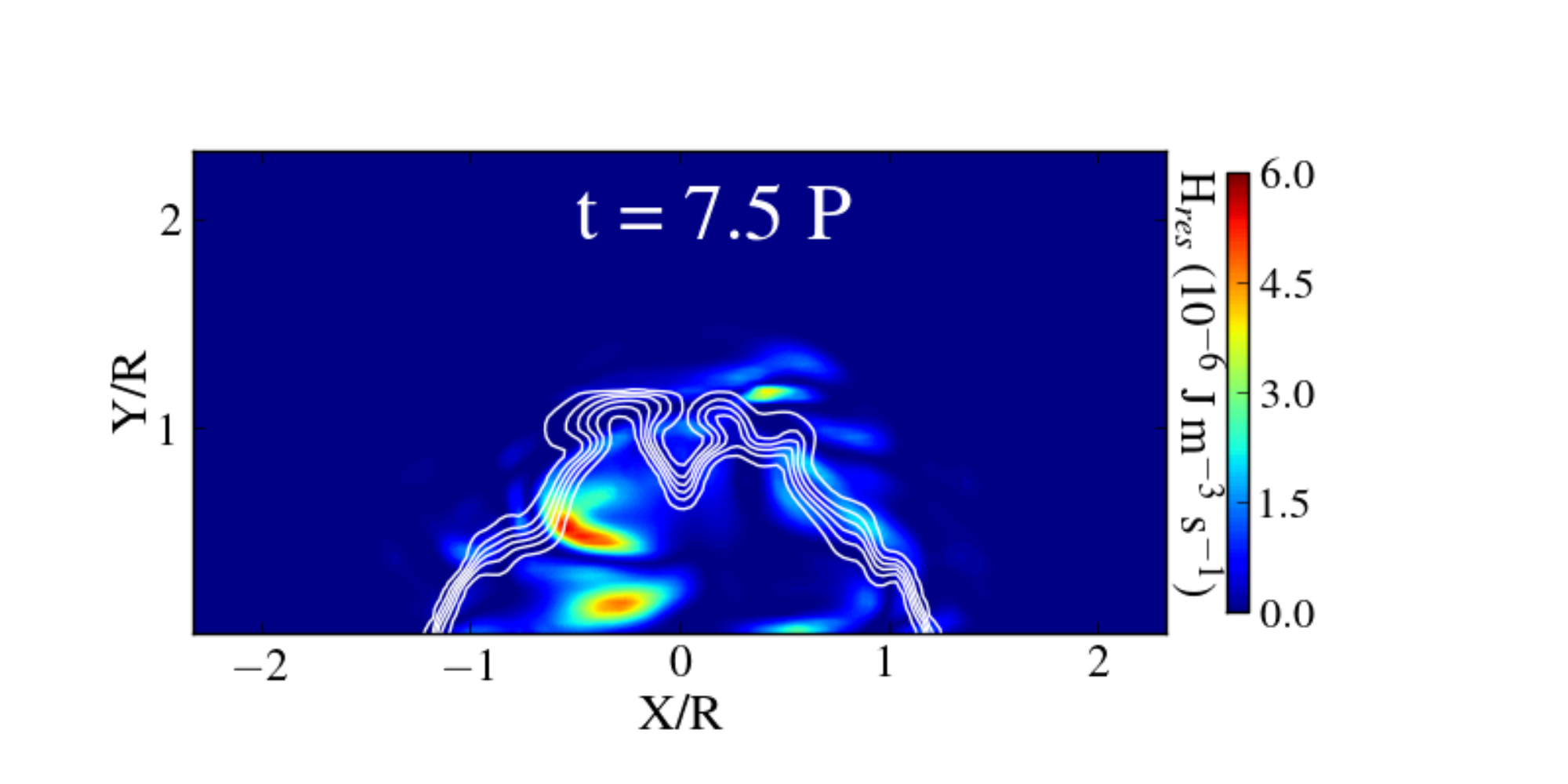}}
\caption{Snapshots of the vorticity $\omega_z^2$ at the apex (top) and of the resistive heating rate $H_{res}$ near the footpoint (bottom), for the driver with $v_0 = 2$ km/s. $P\simeq 254$ s is the driver period. The white lines on both panels represent the density contours at the corresponding heights, showing the circumference of the dense loop.}\label{fig:vortHJ}
\end{figure}

The deformation of the loop cross section due to the KHI instability affects the spatial distribution of the resistive and viscous heating rate. Despite the smooth appearance of the density and temperature in Fig. \ref{fig:struct}, the damping of the kink wave continues to take place. The resonant layer is now turbulently fragmented into many current sheets or shear layers in the velocity, because of the KHI eddies over the whole cross section. The resistive heating peaks near the footpoints, as we see in Fig. \ref{fig:struct}. $H_{res}$, which is dominated by the diffusion of the $J_z$ current density \citep{karampelas2017}, is initially concentrated in the resonant layer. Later, it spreads over the whole tube cross section (Fig. \ref{fig:vortHJ}), even if the loop is not highly deformed near the footpoint. Instead, this is due to the deformed cross section near the loop top affecting the imprint of the current near the footpoint. Likewise, the viscous heating rate is spreading through the entire cross section because of the turbulent deformation (Fig. \ref{fig:vortHJ}). However, in contrast to the resistive heating rate, it finds its maximum near the loop apex.

In our current experiments, there is not enough energy input at the footpoint to balance the energy losses which are to be expected from the plasma (by e.g. optically thin radiation or heat conduction). However, in principle the plasma could be heated if a sufficient energy flux is provided by the driven boundary. The key point of this paper is that this heating (be it resistive or viscous) can take place in the entire cross section of the loop, because of its fully deformed nature. While it was earlier refuted by using drivers with a broad-band spectrum, the decade-old argument \citep{ofman1998,cargill2016ApJ} that wave heating can only take place in specific layers in the loop is thus not even true for monoperiodic drivers.

Despite the fully deformed state of the loop cross section, the forward modelling of our simulation (Fig. \ref{fomotube}) still maintains a loop-like appearance. The only qualitative changes compared to the forward model of a `laminar' loop (i.e. a straight cylinder) is that (1) strand-like features are formed \citep[as previously pointed out by][]{antolin2014fine}, and (2) the overall intensity is increased and spread over a larger layer. The latter is because of the adiabatic expansion during the mixing of the interior and exterior plasma \citep[as previously shown by][]{antolin2016,karampelas2017}. However, the coronal loop structure is clearly distinguishable from its surrounding plasma, and its oscillation remains visible at later times. Thus, the observational identity of a fully deformed loop remains intact, and should be detectable when studying relevant phenomena, such as the decayless loop oscillations from \citet{anfinogentov2015}.

The fact that the heating by the transverse waves occurs in the entire, deformed loop cross section, provides impetus to more detailed wave heating models, going beyond the qualitative arguments presented in this paper. The introduction of a realistic atmosphere and thermal conduction in future setups, the inclusion of physical dissipation terms, such as anomalous resistivity, and the use of stronger drivers could provide a viable loop heated by transverse waves in its entirely deformed cross section, further addressing the issues brought up by \citet{cargill2016ApJ}.

\begin{small}
\textit{Acknowledgements.} We would like to thank the referee, whose review helped us improve the manuscript. We also thank the editor, for his comments. K.K. was funded by GOA-2015-014 (KU Leuven). T.V.D. was supported by the IAP P7/08 CHARM (Belspo) and the GOA-2015-014 (KU Leuven). This project has received funding from the European Research Council (ERC) under the European Union's Horizon 2020 research and innovation programme (grant agreement No 724326). The results were inspired by discussions at the ISSI-Bern and at ISSI-Beijing meetings. 
\end{small}

\bibliographystyle{aa}
\bibliography{aa31646-17}

\end{document}